\documentclass[aps,prl,twocolumn,superscriptaddress,notitlepage]{revtex4-1}
\usepackage{graphicx} 
\usepackage{epsfig}
\usepackage{epstopdf}
\usepackage{amsmath}
\usepackage{amsfonts}
\usepackage{amssymb} 
\usepackage{color}
\usepackage{multirow}

\begin{document}


\title{Rosenbluth separation of the $\pi^0$ electroproduction cross section}
\author{M.~Defurne}
\email[Corresponding author: ]{maxime.defurne@cea.fr}
\affiliation{Irfu, CEA, Universit\'{e} Paris-Saclay, 91191 Gif-sur-Yvette, France}
\author{M.~Mazouz}
\affiliation{Facult\'e des sciences de Monastir, 5000 Tunisia}
\author{H.~Albataineh}
\affiliation{Texas A\&M University-Kingsville, Kingsville, Texas 78363, USA}
\author{K.~Allada}
\affiliation{Massachusetts Institute of Technology,Cambridge, Massachusetts 02139, USA}
\author{K.~A.~Aniol}
\affiliation{California State University, Los Angeles, Los Angeles, California 90032, USA}
\author{V.~Bellini}
\affiliation{INFN/Sezione di Catania, 95125 Catania, Italy}
\author{M.~Benali}
\affiliation{Clermont universit\'{e}, universit\'{e} Blaise Pascal, CNRS/IN2P3, Laboratoire de physique corpusculaire, FR-63000 Clermont-Ferrand, France}
\author{W.~Boeglin}
\affiliation{Florida International University, Miami, Florida 33199, USA}
\author{P.~Bertin}
\affiliation{Clermont universit\'{e}, universit\'{e} Blaise Pascal, CNRS/IN2P3, Laboratoire de physique corpusculaire, FR-63000 Clermont-Ferrand, France}
\affiliation{Thomas Jefferson National Accelerator Facility, Newport News, Virginia 23606, USA}
\author{M.~Brossard}
\affiliation{Clermont universit\'{e}, universit\'{e} Blaise Pascal, CNRS/IN2P3, Laboratoire de physique corpusculaire, FR-63000 Clermont-Ferrand, France}
\author{A.~Camsonne}
\affiliation{Thomas Jefferson National Accelerator Facility, Newport News, Virginia 23606, USA}
\author{M.~Canan}
\affiliation{Old Dominion University, Norfolk, Virginia 23529, USA}
\author{S.~Chandavar}
\affiliation{Ohio University, Athens, Ohio 45701, USA}
\author{C.~Chen}
\affiliation{Hampton University, Hampton, Virginia 23668, USA}
\author{J.-P.~Chen}
\affiliation{Thomas Jefferson National Accelerator Facility, Newport News, Virginia 23606, USA}
\author{C.W.~de~Jager}
\affiliation{Thomas Jefferson National Accelerator Facility, Newport News, Virginia 23606, USA}
\author{R.~de~Leo}
\affiliation{Universit\`{a} di Bari, 70121 Bari, Italy}
\author{C.~Desnault}
\affiliation{Institut de Physique Nucl\'eaire CNRS-IN2P3, Orsay, France}
\author{A.~Deur}
\affiliation{Thomas Jefferson National Accelerator Facility, Newport News, Virginia 23606, USA}
\author{L.~El~Fassi}
\affiliation{Rutgers, The State University of New Jersey, Piscataway, New Jersey 08854, USA}
\author{R.~Ent}
\affiliation{Thomas Jefferson National Accelerator Facility, Newport News, Virginia 23606, USA}
\author{D.~Flay}
\affiliation{Temple University, Philadelphia, Pennsylvania 19122, USA}
\author{M.~Friend}
\affiliation{Carnegie Mellon University, Pittsburgh, Pennsylvania 15213, USA}
\author{E.~Fuchey}
\affiliation{Clermont universit\'{e}, universit\'{e} Blaise Pascal, CNRS/IN2P3, Laboratoire de physique corpusculaire, FR-63000 Clermont-Ferrand, France}
\author{S.~Frullani}
\affiliation{INFN/Sezione Sanit\`{a}, 00161 Roma, Italy}
\author{F.~Garibaldi}
\affiliation{INFN/Sezione Sanit\`{a}, 00161 Roma, Italy}
\author{D.~Gaskell}
\affiliation{Thomas Jefferson National Accelerator Facility, Newport News, Virginia 23606, USA}
\author{A.~Giusa}
\affiliation{INFN/Sezione di Catania, 95125 Catania, Italy}
\author{O.~Glamazdin}
\affiliation{Kharkov Institute of Physics and Technology, Kharkov 61108, Ukraine}
\author{S.~Golge}
\affiliation{North Carolina Central University, Durham, North Carolina 27701, USA}
\author{J.~Gomez}
\affiliation{Thomas Jefferson National Accelerator Facility, Newport News, Virginia 23606, USA}
\author{O.~Hansen}
\affiliation{Thomas Jefferson National Accelerator Facility, Newport News, Virginia 23606, USA}
\author{D.~Higinbotham}
\affiliation{Thomas Jefferson National Accelerator Facility, Newport News, Virginia 23606, USA}
\author{T.~Holmstrom}
\affiliation{Longwood University, Farmville, Virginia 23909, USA}
\author{T.~Horn}
\affiliation{The Catholic University of America, Washington, DC 20064, USA}
\author{J.~Huang}
\affiliation{Massachusetts Institute of Technology,Cambridge, Massachusetts 02139, USA}
\author{M.~Huang}
\affiliation{Duke University, Durham, North Carolina 27708, USA}
\author{C.E.~Hyde}
\affiliation{Old Dominion University, Norfolk, Virginia 23529, USA}
\affiliation{Clermont universit\'{e}, universit\'{e} Blaise Pascal, CNRS/IN2P3, Laboratoire de physique corpusculaire, FR-63000 Clermont-Ferrand, France}
\author{S.~Iqbal}
\affiliation{California State University, Los Angeles, Los Angeles, California 90032, USA}
\author{F.~Itard}
\affiliation{Clermont universit\'{e}, universit\'{e} Blaise Pascal, CNRS/IN2P3, Laboratoire de physique corpusculaire, FR-63000 Clermont-Ferrand, France}
\author{H.~Kang}
\affiliation{Seoul National University, Seoul, South Korea}
\author{A.~Kelleher}
\affiliation{College of William and Mary, Williamsburg, Virginia 23187, USA}
\author{C.~Keppel}
\affiliation{Thomas Jefferson National Accelerator Facility, Newport News, Virginia 23606, USA}
\author{S.~Koirala}
\affiliation{Old Dominion University, Norfolk, Virginia 23529, USA}
\author{I.~Korover}
\affiliation{Tel Aviv University, Tel Aviv 69978, Israel}
\author{J.J.~LeRose}
\affiliation{Thomas Jefferson National Accelerator Facility, Newport News, Virginia 23606, USA}
\author{R.~Lindgren}
\affiliation{University of Virginia, Charlottesville, Virginia 22904, USA}
\author{E.~Long}
\affiliation{Kent State University, Kent, Ohio 44242, USA}
\author{M.~Magne}
\affiliation{Clermont universit\'{e}, universit\'{e} Blaise Pascal, CNRS/IN2P3, Laboratoire de physique corpusculaire, FR-63000 Clermont-Ferrand, France}
\author{J.~Mammei}
\affiliation{University of Massachusetts, Amherst, Massachusetts 01003, USA}
\author{D.J.~Margaziotis}
\affiliation{California State University, Los Angeles, Los Angeles, California 90032, USA}
\author{P.~Markowitz}
\affiliation{Florida International University, Miami, Florida 33199, USA}
\author{A.~Mart\'i Jim\'enez-Arg\"uello}
\affiliation{Facultad de F\'isica, Universidad de Valencia, Valencia, Spain}
\affiliation{Institut de Physique Nucl\'eaire CNRS-IN2P3, Orsay, France}
\author{F.~Meddi}
\affiliation{INFN/Sezione Sanit\`{a}, 00161 Roma, Italy}
\author{D.~Meekins}
\affiliation{Thomas Jefferson National Accelerator Facility, Newport News, Virginia 23606, USA}
\author{R.~Michaels}
\affiliation{Thomas Jefferson National Accelerator Facility, Newport News, Virginia 23606, USA}
\author{M.~Mihovilovic}
\affiliation{University of Ljubljana, 1000 Ljubljana, Slovenia}
\author{C.~Mu\~noz~Camacho}
\affiliation{Clermont universit\'{e}, universit\'{e} Blaise Pascal, CNRS/IN2P3, Laboratoire de physique corpusculaire, FR-63000 Clermont-Ferrand, France}
\affiliation{Institut de Physique Nucl\'eaire CNRS-IN2P3, Orsay, France}
\author{P.~Nadel-Turonski}
\affiliation{Thomas Jefferson National Accelerator Facility, Newport News, Virginia 23606, USA}
\author{N.~Nuruzzaman}
\affiliation{Hampton University, Hampton, Virginia 23668, USA}
\author{R.~Paremuzyan}
\affiliation{Institut de Physique Nucl\'eaire CNRS-IN2P3, Orsay, France}
\author{A.~Puckett}
\affiliation{Los Alamos National Laboratory, Los Alamos, New Mexico 87545, USA}
\author{V.~Punjabi}
\affiliation{Norfolk State University, Norfolk, Virginia 23529, USA}
\author{Y.~Qiang}
\affiliation{Thomas Jefferson National Accelerator Facility, Newport News, Virginia 23606, USA}
\author{A.~Rakhman}
\affiliation{Syracuse University, Syracuse, New York 13244, USA}
\author{M.N.H.~Rashad}
\affiliation{Old Dominion University, Norfolk, Virginia 23529, USA}
\author{S.~Riordan}
\affiliation{Stony Brook University, Stony Brook, New York 11794, USA}
\author{J.~Roche}
\affiliation{Ohio University, Athens, Ohio 45701, USA}
\author{G.~Russo}
\affiliation{INFN/Sezione di Catania, 95125 Catania, Italy}
\author{F.~Sabati\'e}
\affiliation{Irfu, CEA, Universit\'{e} Paris-Saclay, 91191 Gif-sur-Yvette, France}
\author{K.~Saenboonruang}
\affiliation{University of Virginia, Charlottesville, Virginia 22904, USA}
\affiliation{Kasetsart University, Chatuchak, Bangkok, 10900, Thailand}
\author{A.~Saha}
\thanks{Deceased}
\affiliation{Thomas Jefferson National Accelerator Facility, Newport News, Virginia 23606, USA}
\author{B.~Sawatzky}
\affiliation{Thomas Jefferson National Accelerator Facility, Newport News, Virginia 23606, USA}
\affiliation{Temple University, Philadelphia, Pennsylvania 19122, USA}
\author{L.~Selvy}
\affiliation{Kent State University, Kent, Ohio 44242, USA}
\author{A.~Shahinyan}
\affiliation{Yerevan Physics Institute, Yerevan 375036, Armenia}
\author{S.~Sirca}
\affiliation{University of Ljubljana, 1000 Ljubljana, Slovenia}
\author{P.~Solvignon}
\thanks{Deceased}
\affiliation{Thomas Jefferson National Accelerator Facility, Newport News, Virginia 23606, USA}
\author{M.L.~Sperduto}
\affiliation{INFN/Sezione di Catania, 95125 Catania, Italy}
\author{R.~Subedi}
\affiliation{Georges Washington University, Washington, DC 20052, USA}
\author{V.~Sulkosky}
\affiliation{Massachusetts Institute of Technology,Cambridge, Massachusetts 02139, USA}
\author{C.~Sutera}
\affiliation{INFN/Sezione di Catania, 95125 Catania, Italy}
\author{W.A.~Tobias}
\affiliation{University of Virginia, Charlottesville, Virginia 22904, USA}
\author{G.M.~Urciuoli}
\affiliation{INFN/Sezione di Roma, 00185 Roma, Italy}
\author{D.~Wang}
\affiliation{University of Virginia, Charlottesville, Virginia 22904, USA}
\author{B.~Wojtsekhowski}
\affiliation{Thomas Jefferson National Accelerator Facility, Newport News, Virginia 23606, USA}
\author{H.~Yao}
\affiliation{Temple University, Philadelphia, Pennsylvania 19122, USA}
\author{Z.~Ye}
\affiliation{University of Virginia, Charlottesville, Virginia 22904, USA}
\author{A.~Zafar}
\affiliation{Syracuse University, Syracuse, New York 13244, USA}
\author{X.~Zhan}
\affiliation{Argonne National Laboratory, Lemont, Illinois 60439, USA}
\author{J.~Zhang}
\affiliation{Thomas Jefferson National Accelerator Facility, Newport News, Virginia 23606, USA}
\author{B.~Zhao}
\affiliation{College of William and Mary, Williamsburg, Virginia 23187, USA}
\author{Z.~Zhao}
\affiliation{University of Virginia, Charlottesville, Virginia 22904, USA}
\author{X.~Zheng}
\affiliation{University of Virginia, Charlottesville, Virginia 22904, USA}
\author{P.~Zhu}
\affiliation{University of Virginia, Charlottesville, Virginia 22904, USA}
\collaboration{The Jefferson Lab Hall A Collaboration}

\date{\today}

\begin{abstract}
We present deeply virtual $\pi^0$ electroproduction cross-section measurements at $x_B$=0.36 and three different $Q^2$--values ranging from 1.5 to 2~GeV$^2$, obtained from experiment E07-007 that ran in the Hall A at Jefferson Lab. The Rosenbluth technique was used to separate the longitudinal and transverse responses. Results demonstrate that the cross section is dominated by its transverse component, and thus is far from the asymptotic limit predicted by perturbative Quantum Chromodynamics. An indication of a non-zero longitudinal contribution is provided by the interference term $\sigma_{LT}$ also measured. Results are compared with several models based on the leading twist approach of Generalized Parton Distributions (GPDs). In particular, a fair agreement is obtained with models where the scattering amplitude is described by a convolution of chiral-odd (transversity) GPDs of the nucleon with the twist-3 pion distribution amplitude. Therefore, neutral pion electroproduction may offer the exciting possibility of accessing transversity GPDs through experiment.
\end{abstract}


\maketitle

\begin{figure}
\begin{minipage}{0.65\linewidth}
    \includegraphics[width=\linewidth]{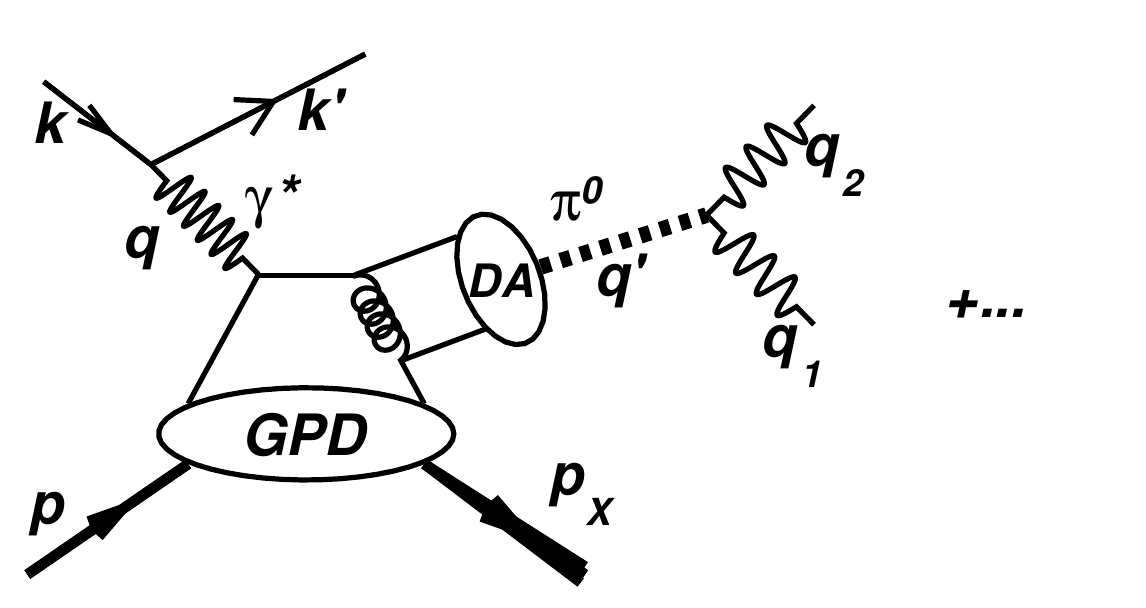}
    \end{minipage}\hfill\begin{minipage}{0.34\linewidth}
    \centerline{Invariants}
    \vskip -1.5em
    \begin{align*}
    Q^2 &= -(k-k')^2 \\
    x_B &= Q^2/(2q\cdot P) \\
    W^2 &= (q+P)^2 \\
    t      &= (q-q')^2  \\
    \end{align*}
    \end{minipage}
    \caption{Diagram of the exclusive $\pi^0$ electroproduction reaction, identified by the  $\pi^0\to \gamma\gamma$ decay mode.
    The value of $t$ with minimal $|t|$ can be evaluated as 
    $t_\text{min} = (Q^2+m_{\pi}^2)^2/(4W^2)-(|q^{\rm c.m.}|-|q^{\prime      
CM}|)^2$,
     with $|q^{\rm c.m.}|$ and $|q^{\prime CM}|$ the norms of $\vec{q}$,        
$\vec{q^{\prime}}$ in the
     $p\pi^0$ final state center-of-mass frame.}
    \label{pi0Diag}
  \end{figure}

Deep exclusive reactions have been the subject of intense experimental and theoretical work in the last decades, as they provide clean probes of the internal three-dimensional structure of hadrons. We present here measurements of the  differential cross section for the forward exclusive electroproduction reaction $ep\to ep\pi^0$. These results are the first separation of the differential cross section for longitudinally and transversely polarized virtual photons of exclusive $\pi^0$-electroproduction in the electron scattering kinematics of Deep Inelastic Scattering (DIS).
A diagram of this process, including definitions of the kinematic variables,
is presented in Fig.~\ref{pi0Diag}.

 The Quantum Chromodynamics (QCD) factorization theorems predict that deep virtual meson production should be dominated by the longitudinal virtual photo-production cross section~\cite{Collins:1996fb}. In the Bjorken limit $Q^2\to\infty$ and $t/Q^2\ll 1$ at fixed $x_B$, the longitudinal scattering amplitude factorizes into a hard perturbative contribution, the leading twist Generalized Parton Distributions (GPDs) of the nucleon and the pion distribution amplitude (DA)~\cite{Collins:1996fb,Vanderhaeghen:1999xj, Goeke:2001tz}. GPDs are light-cone matrix elements of non-local bilinear quark and gluon operators that describe the three-dimensional structure of hadrons, by correlating the internal transverse position of partons to their longitudinal momentum~\cite{Mueller:1998fv,Ji:1996ek,Radyushkin:1997ki}. In the case of a nucleon, 8 GPDs describe at leading twist the different combinations of parton and nucleon helicities. Four chiral-even GPDs conserve the helicity of the parton, whereas four chiral-odd GPDs, also referred to as transversity GPDs, flip its helicity. While a rigorous factorization proof has not been established for the transverse virtual photo-production amplitude, it is proven to be suppressed by a factor of $1/Q$ with respect to its longitudinal counterpart~\cite{Collins:1996fb}.

The leading-twist approximation is in good agreement with  high $Q^2$ electroproduction data for photon~\cite{Airapetian:2008aa,Defurne:2015kxq,Jo:2015ema} and vector meson production~\cite{Wolf:2009jm, Favart:2015umi}. On the other hand, the co-linear approximation underestimates by about one order of magnitude the total $\pi^0$ electroproduction cross sections measured at $Q^2\sim 2$~GeV$^2$ by the Hall A~\cite{Collaboration:2010kna} and CLAS~\cite{Bedlinskiy:2012be} collaborations at Jefferson Lab (JLab). It was suggested in \cite{Ahmad:2008hp} that for neutral meson production the twist-3 quark-helicity flip pion DAs coupled with the transversity GPDs of the proton would create a large cross section for transversely polarized deeply virtual photons, without violating the QCD factorization theorem.

The Deeply Virtual Meson Production (DVMP) cross section can be written in the following form~\cite{DT}:
\begin{multline}
\frac{d^4\sigma}{dQ^2 dx_B dt d\phi}=\frac{1}{2\pi}\frac{d^2\Gamma}{dx_B dQ^2}(Q^2, x_B, E)\Big[\frac{d\sigma_T}{dt}+\epsilon\frac{d\sigma_L}{dt}\\+\sqrt{2\epsilon (1+\epsilon)}\frac{d\sigma_{TL}}{dt}\cos\phi+\epsilon \frac{d\sigma_{TT}}{dt}\cos 2\phi\Big]\;,
\end{multline}
where $E$ is the incident lepton energy in the target rest frame and $\phi$ the angle between the leptonic and hadronic plane defined according to the Trento convention~\cite{Bacchetta:2004jz}. The factor $\frac{d^2\Gamma}{dx_B dQ^2}(Q^2, x_B, E)$ is the virtual photon flux and $\epsilon$ is the degree of longitudinal polarization defined as ($y=[q\cdot p]/[k\cdot p]$):
\begin{align}
&\frac{d^2\Gamma}{dx_B dQ^2}(Q^2, x_B, E)=\frac{\alpha}{8\pi}\frac{Q^2}{M^2 E^2}\frac{1-x_B}{x_B^3}\frac{1}{1-\epsilon},
\\
&\epsilon=\frac{1-y-{Q^2}/{4E^2}}{1-y+{y^2}/{2}+{Q^2}/{(4E^2)}}\;,
\end{align}
$M$ being the proton mass.

Experiment E07-007 ran in the JLab Hall A from October to December 2010. One of its goals was to separate the exclusive transverse and longitudinal $\pi^0$ electroproduction cross sections using the Rosenbluth technique, consisting on measurements at two different values of the incident electron energy. Tab.~\ref{tab:pi0kin} lists the three $Q^2$ settings measured, each of them at two different values of $\epsilon$. 
\begin{table}
\caption{
E07-007 $ep\rightarrow e p \pi^0$ kinematic settings
}
\begin{ruledtabular}
\begin{tabular}{ccccc}
Setting & $Q^2$ & $x_B$& $E^{beam}$ & $\epsilon$\\
& (GeV$^2$)&& (GeV) &\\\hline
\multirow{2}{*}{Kin1} &  \multirow{2}{*}{1.50} & \multirow{2}{*}{0.36} & 3.355 & 0.52 \\ \cline{4-5}
&&& 5.55 & 0.84 \\ \hline
\multirow{2}{*}{Kin2}&\multirow{2}{*}{1.75} & \multirow{2}{*}{0.36}& 4.455 & 0.65 \\ \cline{4-5}
&&& 5.55 & 0.79 \\ \hline
\multirow{2}{*}{Kin3}&\multirow{2}{*}{2.00}& \multirow{2}{*}{0.36}&  4.455 & 0.53 \\ \cline{4-5}
&&& 5.55 & 0.72 \\ 
\end{tabular}
\end{ruledtabular}
\label{tab:pi0kin}
\end{table}

The electron beam was incident on a 15-cm-long liquid H$_2$ target, for a typical luminosity of 2$\cdot$10$^{37}$~cm$^{-2}$s$^{-1}$. Scattered electrons were detected in a high resolution spectrometer (HRS), with $\sim10^{-4}$ momentum resolution and $<2$~mr horizontal angular resolution~\cite{Alcorn:2004sb}. The two photons of the $\pi^0$ decays were detected in a PbF$_2$ electromagnetic calorimeter consisting of a 13$\times$16 array of 3 $\times$ 3 $\times$ 18.6~cm$^3$ crystals, coupled to mesh-dynode photomultipliers. Each calorimeter channel was continuously sampled by a 1~GHz flash ADC system that recorded the signal over 128~ns for every event. The high resolution in the electron kinematics determined accurately the values of $Q^2$ and $x_B$. The fast \v{C}erenkov signal from the calorimeter allowed a coincident time resolution between the electron and $\pi^0$ detections of 0.6~ns. The vertex resolution of the HRS and position resolution of the calorimeter accurately determined the $\pi^0$ direction and thus the kinematical variables $t$ and $\phi$. The measured energy in the calorimeter is used to identify $\pi^0$ events through the 2-photon invariant mass $m_{\gamma\gamma}=\sqrt{(q_1+q_2)^2}$ and to ensure the exclusivity of the reaction using the $ep\to e\gamma \gamma X$ missing mass squared $M^2_{ep\to e\gamma \gamma X}$.

The calibration of the calorimeter was done in two steps. Firstly, we used elastic scattering $H(e,e_\text{Calo}^\prime p_\text{HRS})$ events. This calibration required dedicated runs, since the polarity of the HRS had to be reversed to allow proton detection. We performed elastic calibrations at the beginning, middle and end of the experiment. A resolution of 3.1\% at 3.16 GeV was measured, with a position resolution of 3~mm at 110~cm from the target. Between elastic calibrations, channel gains were observed to drift up to 10\%. We attributed these changes to radiation damage of the PbF$_2$ crystals. In order to correct for the calibration coefficient drifts between the elastic run periods we used exclusive $\pi^0$ data from our $H(e,e'\gamma\gamma)X$ sample. By assuming $M_{ep\to e\gamma \gamma X}^2=M^2$ and $m_{\gamma \gamma}$=$m_{\pi^0}$, the sum of the energies of the two decay photons was determined and used to compute the calibration coefficients. The combination of both elastic and exclusive $\pi^0$ calibrations provided a continuous invariant mass resolution of 9.5~MeV through the full run period.

\begin{figure}[!hbt]
\centering 
\includegraphics[width=\linewidth]{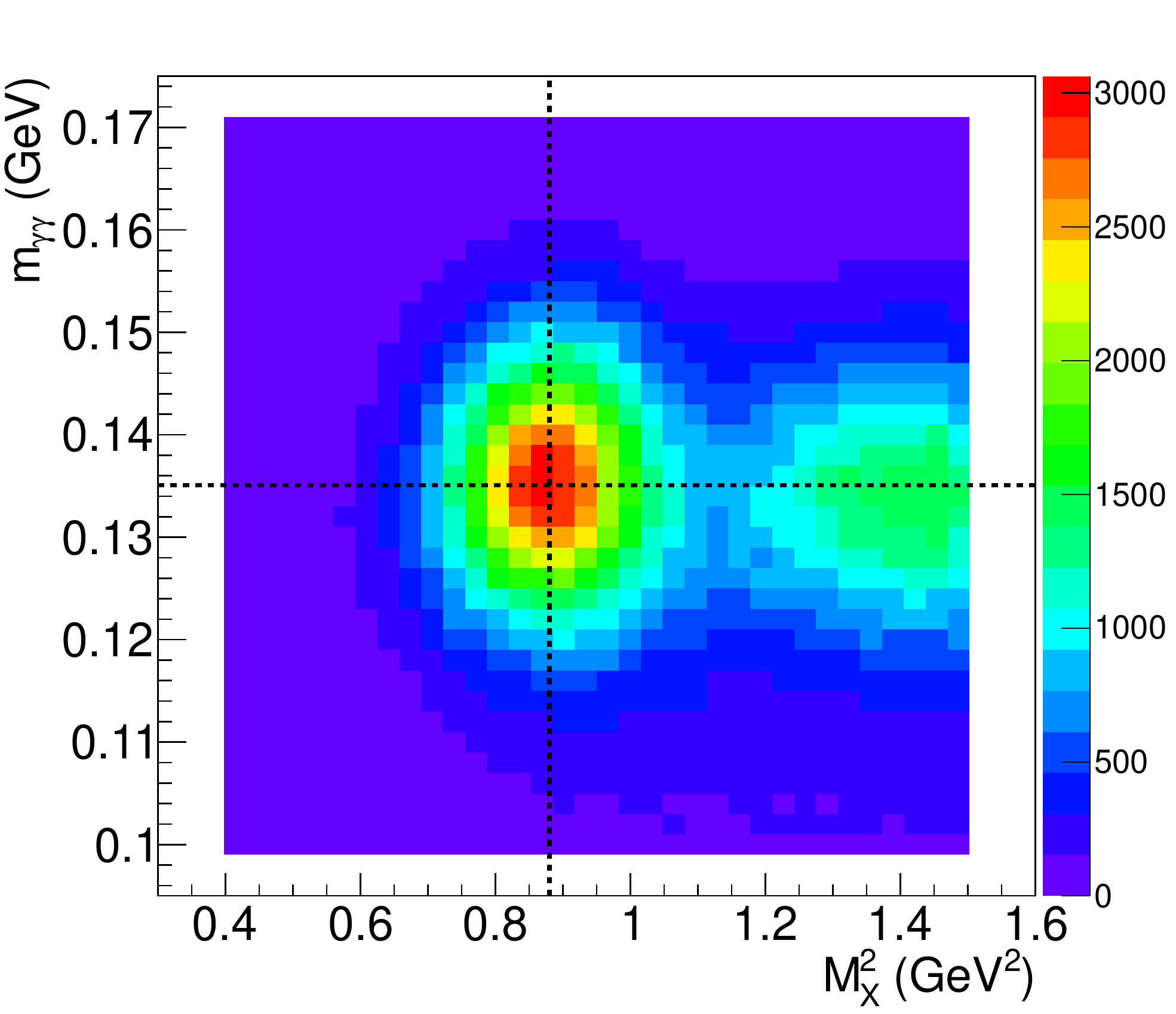}
\caption{\label{mggM2}(Color online) Distributions of the $H(e,e'\gamma\gamma)X$ events within cuts in the [$M_{X}^2,m_{\gamma\gamma}$] plane. Dotted lines illustrate the $\pi^0$ mass and the
proton mass squared.}
\end{figure}

The data acquisition was triggered by an electron detection signal in the HRS, formed by the coincidence of the gas \v{C}erenkov detector and the plastic scintillator plane S$2$m of the HRS~\cite{Alcorn:2004sb}. In order to select neutral pions, we studied 2-cluster events in the calorimeter with an energy deposit $\ge 500$~MeV in each cluster and within 3~ns of the electron detection. To account for the natural correlation between $M_{ep\rightarrow e \gamma \gamma X}^2$ and $m_{\gamma \gamma}$, we define a correction such as:
\begin{equation}
M^2_X=M^2_{ep\rightarrow e \gamma \gamma X}+C\times \left(m_{\gamma \gamma}-m_{\pi^0}\right)\;,
\end{equation}
with the empirical value $C=12$~GeV. Fig.~\ref{mggM2} shows the distribution of the $H(e,e'\gamma\gamma)X$ events in the [$M_{X}^2,m_{\gamma\gamma}$] plane, which exhibits a clean signal centered on the pion mass and proton mass squared. Exclusive events are selected by requiring $100<m_{\gamma \gamma}<170$~MeV and $M^2_X<0.95$~GeV$^2$. The number of accidental $H(e,e'\gamma\gamma)X$ triple coincidences is estimated by measuring the number of 2-photon events detected in the calorimeter for each of the three possible timings with respect to the scattered electron: one photon in-time and one out-of-time, both out-of-time but in-time between themselves, and both out-of-time with the electron and with each other. Finally, an analysis of 3-cluster events was performed in order to correct for the fraction of exclusive $\pi^0$ events where one of the 3 clusters was an accidental photon coincidence. This correction was applied bin-by-bin and found to be 5\% on average.

The different terms of the unpolarized $\pi^0$ cross section are extracted by minimizing the following $\chi^2$ defined between the number of experimental and simulated events:
\begin{equation}
\label{eq::fit}
\chi^2=\sum_{i=0}^{N}\left(\frac{N^{exp}_i-N^{sim}_i}{\sigma^{exp}_i}\right)^2\;,
\end{equation}
where the sum runs over all experimental bins of one $Q^2$ setting, including data at two different values of $\epsilon$. The variable $N^{exp}_i$ is the number of events in the experimental bin $i$, with $\sigma^{exp}_i$ being its corresponding uncertainty. The number of simulated events $N^{sim}_i$ is given by: 
\begin{equation}
\label{eq::nsim}
N^{sim}_i=\mathcal{L} \int_i \frac{d\sigma}{dt dQ^2 dx_B d\phi} dt dQ^2 dx_B d\phi\;,
\end{equation}
with $\mathcal{L}$ the experimental integrated luminosity, corrected by the data acquisition dead-time. The integration is performed with a Monte-Carlo simulation, convoluting the known kinematical dependences of the cross section with the experimental acceptance. We limit the analysis to the overlapping ($Q^2$;$x_B$)-phase space between the two beam-energy settings. After minimization of Eq.~\eqref{eq::fit}, the unknown $Q^2$--dependences of ${d\sigma_T}/{dt}$,${d\sigma_L}/{dt}$, ${d\sigma_{TT}}/{dt}$ and ${d\sigma_{TL}}/{dt}$ are fitted to the results and included into the Monte-Carlo simulation in order to account for the leading variations of the cross section within bins. A second $\chi^2$ minimization is performed, which provides stable results over further iterations and yield the final results we present herein, with $\chi^2/dof$=75.8/59, 82.9/79 and 60.9/59 respectively for $Q^2$=1.50, 1.75 and 2.00~GeV$^2$. No $Q^2$--dependence is included for ${d\sigma_L}/{dt}$ as results are found compatible with zero in all experimental bins. Tab.~\ref{tab:Qdep} shows the $Q^2$--dependences obtained. The small HRS acceptance does not allow for an $x_B$-dependence study.

\begin{table}[t]
\centering
\begin{tabular}{lcr}
\hline
{\textbf {Term}$\qquad$} &  \textbf{$n_{exp}$} & \textbf{$n_{theo}$} \\
\hline
\hline
$d\sigma_T/dt$ & 9 $\pm$ 2 & 8\\
$d\sigma_{TT}/dt$ & 4 $\pm$ 2 & 8\\
$d\sigma_{TL}/dt$ & 26 $\pm$ 5 & 7\\
\hline
\end{tabular}
\caption{$Q-$dependence determined by fitting the $t-$integrated responses with the function $A/Q^{n_{exp}}$. The  QCD asymptotic limit of each term is $\sim Q^{-n_{theo}}$.}
\label{tab:Qdep}
\end{table} 

The Monte-Carlo simulation is based on the GEANT4 toolkit. It includes radiative corrections following the procedure described in~\cite{Defurne:2015kxq} based on calculations by Vanderhaeghen et al.~\cite{Vanderhaeghen:2000ws}. The HRS acceptance is modeled by an R-function that defines the distance of the particle from the HRS acceptance bound~\cite{Rvachev:2001}. Our cut on $M^2_X$ to ensure exclusivity removes a significant fraction of exclusive $\pi^0$ events. This is compensated by applying an identical cut on the simulated data. For this to be accurate, the experimental and Monte-Carlo simulated $M^2_X$ (and $m_{\gamma\gamma}$) distributions should have exactly the same widths and positions. These parameters are dominated by the calibration and resolution of the electromagnetic calorimeter crystals. Thus, great care was taken to locally reproduce the calorimeter energy and position resolutions in the Monte-Carlo simulation. The systematic uncertainty associated to the mismatch between the experimental and simulated $M_X^2$ distributions was estimated to be around 2\%. This uncertainty depends on $\phi$ and $t$ since these variables are strongly correlated to the photon impact point in the calorimeter. In order to propagate it to the extraction of the four structure functions, we added it in quadrature to the statistical uncertainty when computing the $\sigma^{exp}_i$ of each bin in Eq.~\eqref{eq::fit}.

Tab.~\ref{tab:sys} lists the different sources of correlated systematic uncertainties. A check of our global normalization was made by extracting the DIS cross section in each of our kinematic settings. Results agree within the uncertainty listed in Tab.~\ref{tab:sys} with the parametrization of the DIS cross section in~\cite{Accardi:2009br}.

\begin{table}[t]
\centering
\begin{tabular}{lr}
{\textbf {Systematic uncertainty}$\qquad\qquad$} & {\textbf{ Value}} \\
\hline
\hline
HRS acceptance cut & 1\% \\
Gas \v{C}erenkov detector efficiency & 0.5\%  \\
HRS tracking efficiency & 0.5\% \\
$\pi^0$ detection efficiency & 0.5\% \\
Radiative corrections & 2\%  \\
Deadtime and luminosity & 2\% \\
\hline
\hline
Total & 3.12\%\\
\hline
\end{tabular}
\caption{Normalization systematic uncertainties in the extracted $\pi^0$ electroproduction cross sections. They are approximately correlated in $\phi$ and $t$.}
\label{tab:sys}
\end{table} 

\begin{figure}[b!]
\centering
\includegraphics[width=\linewidth]{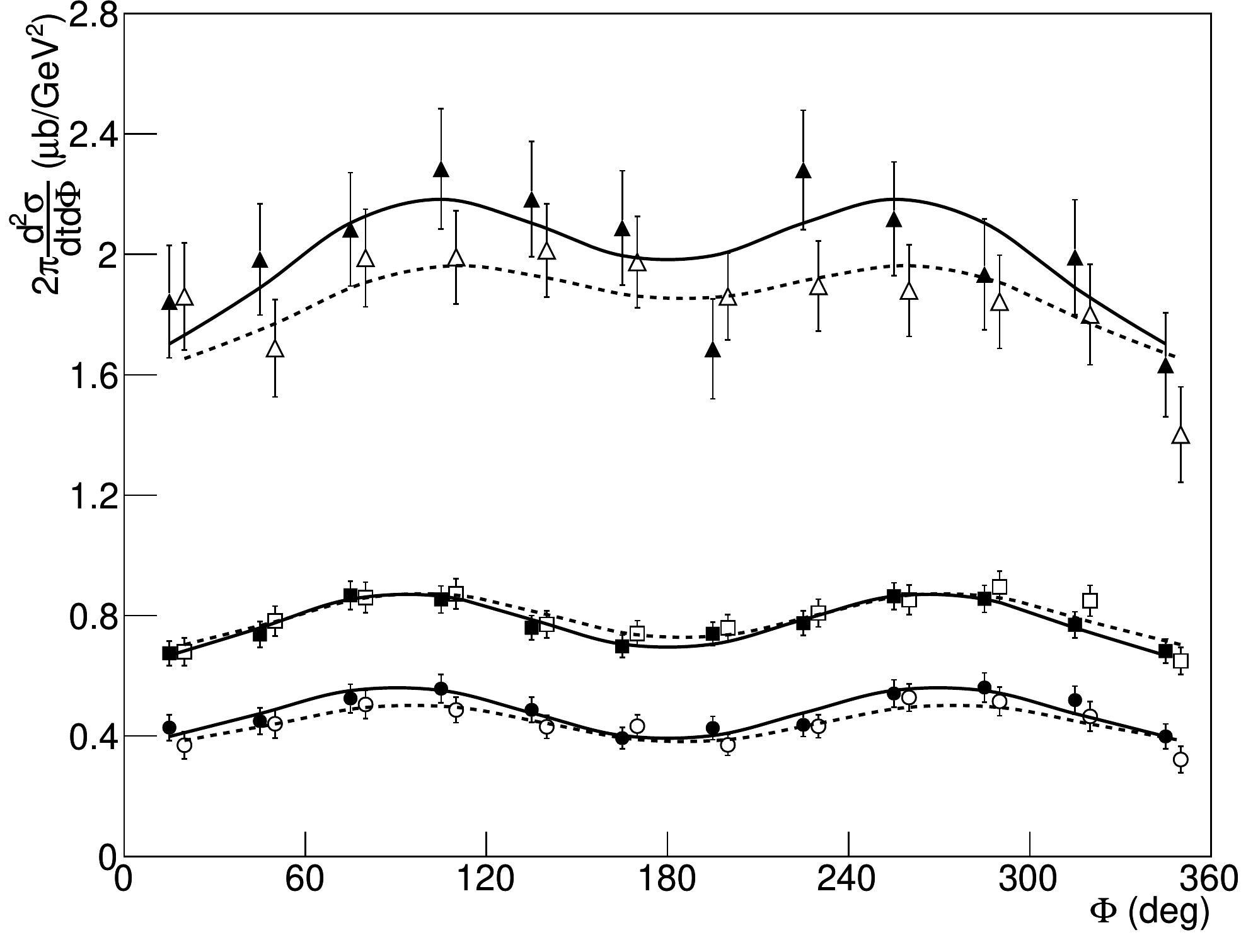}
\caption{\label{fig::2eps}$2\pi\frac{d^2\sigma}{dtd\phi}$ for $Q^2$=1.5 (triangles), 1.75 (squares) and 2~GeV$^2$ (circles) at x$_B$= 0.36 and $t_{min}-t$= 0.025~GeV$^2$. The cross sections extracted at low/high $\epsilon$ are shown in open/filled symbols (and dashed/solid lines).}
\end{figure}

Fig.~\ref{fig::2eps} presents the electroproduction cross section $2\pi\frac{d^2\sigma}{dtd\phi}$ for the three different $Q^2$--values and the lowest $t'=t_{min}-t$ bin, as a function of $\phi$. The cross section is almost independent of $\epsilon$, indicating that most of the signal is coming from its transversely polarized component.   

\begin{figure*}[!hbt]
\centering 
\includegraphics[width=\linewidth]{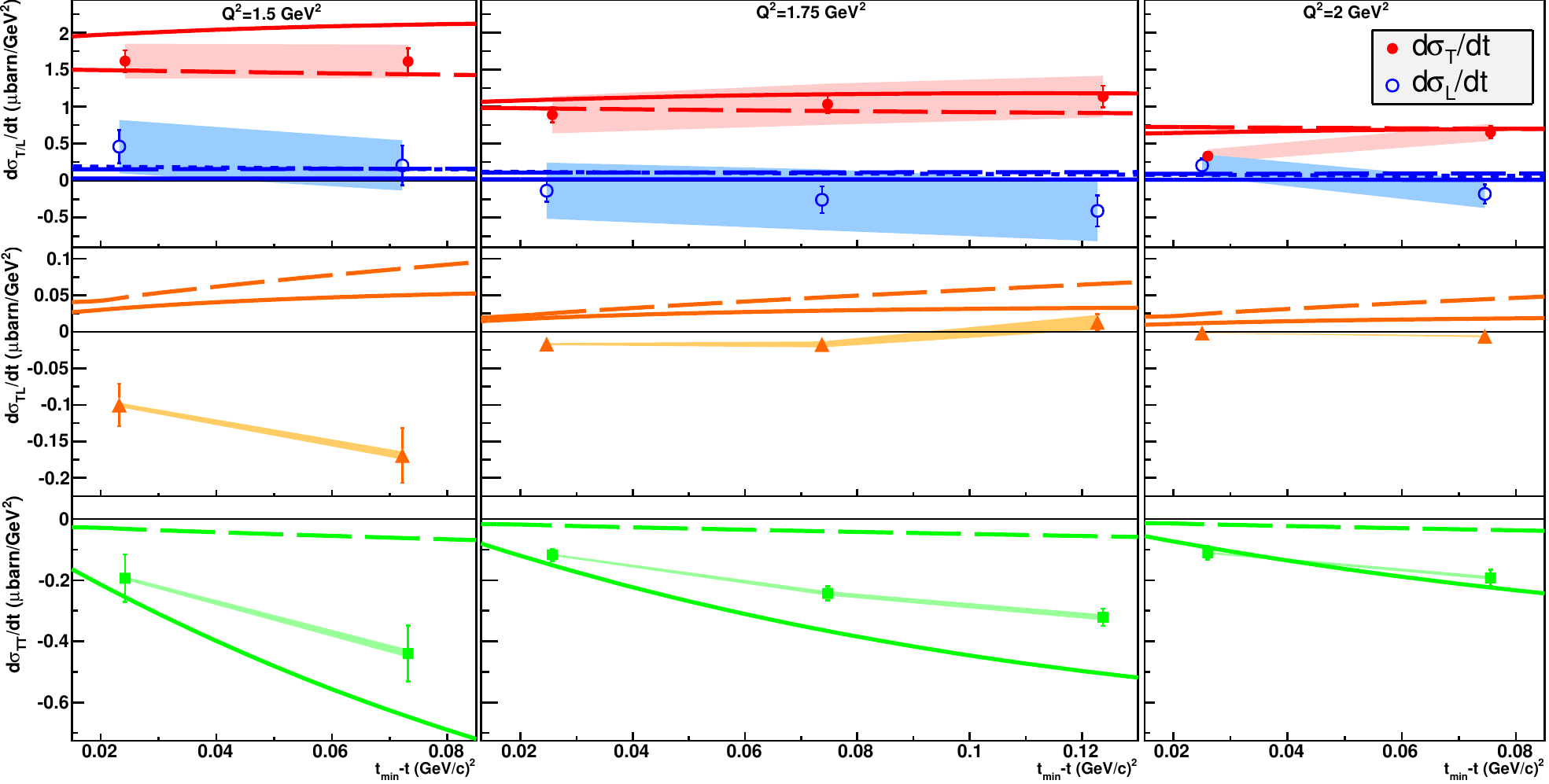}
\caption{\label{fig::LT}(Color online) $\frac{d\sigma_T}{dt}$ (full circles), $\frac{d\sigma_L}{dt}$ (open circles), $\frac{d\sigma_{TL}}{dt}$ (triangles) and $\frac{d\sigma_{TT}}{dt}$ (squares) as a function of $t_{min}-t$ for Q$^2$=1.5 (left), 1.75 (center)  and 2~GeV$^2$ (right) at x$_B$=0.36. The full lines are predictions from the Goloskokov-Kroll model~\cite{Goloskokov:2011rd} and the long-dashed lines from the Liuti-Goldstein model~\cite{Goldstein:2010gu}. The short-dashed line are predictions from the VGG model~\cite{Vanderhaeghen:1999xj} for $\frac{d\sigma_L}{dt}$. Bands connecting the data points show normalization systematic uncertainties on the experimental data; for $d\sigma_L/dt$ and $d\sigma_T/dt$ these bands are strongly anti-correlated}
\end{figure*} 

The uncertainties of the Rosenbluth separated $\frac{d\sigma_L}{dt}$ and $\frac{d\sigma_T}{dt}$ are amplified by the limited lever-arm in $\epsilon$ and the small ratio $\frac{d\sigma_L}{dt}$/$\frac{d\sigma_T}{dt}$. Once the normalization uncertainty is propagated, $\sigma_L$ is found to be compatible with zero, as seen in Fig.~\ref{fig::LT}. However, the interference cross section $\frac{d\sigma_{TL}}{dt}$ is sizeable, which means that $\frac{d\sigma_L}{dt}$, though small, is not negligible. The fact that $\frac{d\sigma_T}{dt}\gg\frac{d\sigma_L}{dt}$ shows that this kinematic regime is still far from the asymptotic prediction of perturbative QCD. These results are compared to several models. The leading twist chiral-even GPD VGG model~\cite{Vanderhaeghen:1999xj} predicts a very small longitudinal cross section, compatible with our results. Two models using a modified factorization approach are also shown in Fig.~\ref{fig::LT}~\cite{Goldstein:2010gu,Goloskokov:2011rd}. In these models leading twist chiral-odd (transversity) GPDs of the nucleon are coupled to a twist-3 DA of the pion, and singularities that prevent collinear factorization in the case of transversely polarized photons are regularized by the transverse momentum $k_\perp$ of the quarks and antiquarks making up the meson. Transversity models are in good agreement with our results of $\frac{d\sigma_T}{dt}$ and $\frac{d\sigma_L}{dt}$ within the experimental uncertainties. However, they predict the opposite sign for $\frac{d\sigma_{TL}}{dt}$ and fail to reproduce the $Q-$dependence of the interference terms listed in Tab.~\ref{tab:Qdep}.

In conclusion, we have performed the L/T separation of $\pi^0$ electroproduction cross section for $Q^2$= 1.5, 1.75 and 2.0~GeV$^2$ at $x_B$=0.36. $\frac{d\sigma_L}{dt}$ is found compatible with zero in all of our experimental bins. The fair agreement between our results and two transversity-GPD models supports the prediction of a chirally enhanced helicity-flip pion distribution amplitude~\cite{Ahmad:2008hp,Goloskokov:2011rd} and the exciting possibility of accessing transversity GPDs of the nucleon through exclusive $\pi^0$ electroproduction. 

We thank G.~Goldstein, S.~Goloskokov, M.~Guidal, P.~Kroll, S.~Liuti and M.~Vanderhaeghen for valuable information about their work and providing the results of their models. We acknowledge essential work of the JLab accelerator staff and the Hall A technical staff. This work was supported by the Department of Energy (DOE), the National Science Foundation, the French {\em Centre National de la Recherche Scientifique}, the {\em Agence Nationale de la Recherche}, the {\em Commissariat \`a l'\'energie atomique et aux \'energies alternatives} and P2IO Laboratory of Excellence. Jefferson Science Associates, LLC, operates Jefferson Lab for the U.S. DOE under U.S. DOE contract DE-AC05-060R23177.

\bibliography{Compton2014}

\end{document}